\documentclass[apj]{emulateapj}
\usepackage{natbib}

\usepackage[utf8]{inputenc}
\usepackage{multirow}
\DeclareUnicodeCharacter{2010}{-}
\usepackage{amsmath}
\usepackage{amsfonts}
\usepackage{amssymb}
\usepackage{todonotes}
\usepackage{ulem}

\begin{document}
	
\title{Spherical shocks in a steep density gradient of expanding media}
\author{Taya Govreen-Segal, Ehud Nakar, Amir Levinson}
\affil{School of Physics \&
Astronomy, Tel Aviv University, Tel Aviv 69978, Israel}

\begin{abstract}
We study the propagation of a Newtonian shock in a spherically symmetric, homologously expanding ejecta. We focus on media with a steep power-law density profile of the form $\rho \propto t^{-3}v^{-\alpha}$, with $\alpha>5$, where $v$ is the velocity
of the expanding medium and $t$ is time. Such profiles are expected in the leading edge of supernovae ejecta and sub-relativistic outflows from binary neutron star mergers.  We find that such shocks always accelerate in the lab frame and lose causal contact with the bulk of the driver gas, owing to the steep density profile. However, the prolonged shock evolution exhibits two distinct pathways: In one, the shock strength diminishes with time until the shock eventually dies out. In the other, the shock strength steadily increases, and the solution approaches the self-similar solution of a shock is a static medium. By mapping the parameter space of shock solutions, we find that the evolutionary pathways are dictated by $\alpha$ and by the initial ratio between the shock velocity and the local upstream velocity. We find that for $\alpha<\omega_c$  ($\omega_c \approx 8$), the shock always decays, and that for $\alpha>\omega_c$ the shock may decay or grow stronger depending on the initial value of the velocity ratio. These two branches bifurcate from a self-similar solution derived analytically for a constant velocity ratio. We analyze properties of the solutions that may have an impact on the observational signatures of such systems, and assess the conditions required for decaying shocks to break out from a finite medium.
\end{abstract}

\section{Introduction}
The strong explosion problem, consisting of a shock driven into an external medium by a sudden release of energy, has been broadly studied. Particular attention has been given to shock propagation in spherically symmetric, power-law density profiles, for which asymptotic self-similar solutions can be found. Solutions of this kind are relevant to diverse phenomena, where the initial conditions have been forgotten, and the symmetry of the problem is close to spherical. Such phenomena include a wide range of explosions and eruptions of stellar systems and are ubiquitous in astrophysics.

In most astrophysical systems the medium into which the shock propagates is at rest. However, there are cases wherein a shock may be driven into an expanding medium.  One example is a class of  various types of super-luminous supernovae (SLSN; for review see \citealt{Gal-Yam2019}), for which a plausible scenario is the injection of energy by a central engine, conceivably a magnetar or accreting black hole, into an expanding supernova remnant \citep[e.g.][]{Kasen2010}. Another example is the merger of a binary neutron star.  Here, a considerable amount of mass is ejected during the merger,
just prior to the formation of a compact engine at the center of the system
 (see \citealt{Nakar2019} for a review). The central engine expels a narrowly collimated, relativistic jet that interacts with the ejecta, 
 and it is possible that it also generates a much wider wind that drives a quasi spherical shock into the ejecta \citep[e.g.,][]{Beloborodov2020}.

The outflow into which the shock is driven in these scenarios expands homologously with a gradient of velocities, that is, every mass element propagates with a constant velocity, $v$, such that at any given radius $r=v t$, where $t$ is the time elapsed since the outflow was ejected.  In the inner slower parts of the outflow the density gradient is relatively shallow, however at the leading front the gradient can be very steep. For instance, in supernovae 
the structure of the ejecta is dictated by the density profile of the pre-supernova static progenitor envelope, which is very steep near the stellar edge, decreasing roughly as a power-law of the distance from the edge.  The emergence of a spherical shock from the static envelope following the SN explosion accelerates the matter near the edge of the envelope to high velocities,  ultimately generating an outflow with a velocity profile
of the form $\rho \propto v^{-\alpha}$ above some characteristic velocity of $0.01-0.1$c (c is the speed of light), with $\alpha \approx 10-12$ for typical stars \citep[e.g.,][]{Nakar2010a}. A second shock driven by a putative central engine that forms following the SN explosion will propagate
in the expanding envelope until reaching the leading edge, whereupon it will start accelerating in the steep density gradient. Another example is the structure of the ejecta from a binary neutron star merger. This structure is not 
well resolved currently, yet we know that the bulk of the ejecta is moving at a velocity of $0.1-0.3$c, and numerical simulations as well as analytic considerations suggest that there is a low-mass fast tail with a sharp velocity gradient that extends to mildly, and possibly even ultra, relativistic velocities \citep[e.g.,][]{Kyutoku2014,Radice2018,Beloborodov2020}. For example, \citet{Hotokezaka2018} find a fast tail with a velocity profile that can be approximated over a limited velocity range as a power law with an index $\alpha \approx 10$.  

The goal of this paper is to study the physics of spherical shocks that propagate in a steep density gradient, such as those expected in the ejecta of the supernovae and binary neutron star mergers. Such shocks can have several interesting observational signatures. First, if the ejecta is optically thin, then the shock generates an observable radiation with properties that reflect the shock velocity and the density gradient. Second, if the ejecta is sufficiently opaque then the shock is radiation mediated (see \citealt{levinson2020} for a review). If the shock is strong enough to cross the ejecta all the way to the point where its optical depth to the observer drops below $c/v$, then the shock breaks out, generating a bright flare \citep[e.g.,][]{Kasen2016} which is followed by emission
of photons that continuously diffuse out of the expand-
ing envelope to the observer (known as cooling emission).
Finally, a shock that traverses the ejecta modifies its den-
sity profile; this profile, in turn, affects the observed emis-
sion through either absorption or emission, since the pho-
tons that ultimately reach the observer diffuse through
the ejecta before escaping the system.Moreover, if the shocked 
ejecta encounters external medium then the generated  emission  also depends, among other things, on the post-shock profile. 

Almost all previous studies of the strong explosion problem have focused on  explosions in static media. The solutions for this problem may be roughly divided into two types, depending on the density gradient in the medium. If the gradient is shallow enough the shock decelerates and  stays in causal contact with the entire region behind it.  In that case, the energy associated with the shock is a conserved quantity of the system. On the other hand, if the gradient is steep enough, the shock accelerates and loses causal contact with part of the mass behind it, and the energy connected to the shock diminishes as the shock propagates. For a power-law density profile in a static medium there exist self similar solutions in the Newtonian and the relativistic regimes for both, decelerating shocks  \citep[e.g.,][]{sedov1946propagation,von1947blast,Taylor1950,Blandford1976} and accelerating 
shocks \citep{Waxman1993,Sari2006}. The solutions for a shock propagating in an expanding medium are fundamentally different since 
the existence of  two characteristic velocities -  that of the ejecta just ahead of the shock, and that of the shock itself - breaks, quite generally, 
the self-similarity of the system.

Previous studies on shock propagation in expanding media have focused on the interaction of a pulsar wind or a magnetar wind with the bulk of the mass of a supernova remnant \citep{Chevalier1984,Jun1998,Suzuki2017}. This is different than the problem considered here since the density gradient is shallow, $\alpha<3$, meaning the pulsar wind continuously injects energy into the region which is causally connected to the shock at all times.  Under these conditions, for self-similar ejecta, there exist  self-similar solutions that describe a double-shock (forward/reverse shock) system.  

In this paper we study a different evolutionary regime. First, we are interested in the propagation of the shock in the steep density gradient tail leading the ejecta. Second, we mainly consider the evolution of a shock driven by a sudden explosion (i.e., instantaneous energy injection into the ejecta), although we show that under a wide range of conditions our solutions are also applicable to a continuous energy injection by a fast wind. To be more specific, we consider the problem of a shock propagating in a homologously expanding, spherically symmetric medium, with a density profile described by a power-law of the form $ \rho_{ej} \propto v^{-\alpha}$, with $\alpha> 5$. We focus on Newtonian shock dynamics, and an ideal equation of state with an adiabatic index, $\gamma$, of $\frac{4}{3}$ and $\frac{5}{3}$. 

As will be discussed later on, the distinction between accelerating and decelerating shocks is not useful in the case considered here, since all the shocks in the regime we study are accelerating in the lab frame.  The key feature that distinguishes between different types of shocks is
the ratio between the shock velocity and the local upstream velocity.  We identify two families; one in which this ratio increases monotonically and one in which it decreases monotonically.   Shocks that belong to the first family have a steadily increasing
strength during their evolution, and are henceforth termed {\it growing shocks}, whereas the evolution of shocks that belong to the second family, henceforth termed 
 {\it decaying shocks}, approaches at some point a phase during which their strength diminishes over time until they completely die out.

Our analysis indicates that for a density profile with $\alpha<\omega_c$ ($\omega_c\simeq 8.22$ for $\gamma=\frac{4}{3}$, and $\omega_c\simeq 7.69$ for $\gamma = \frac{5}{3}$), the shock wave always decays, while for a steeper density profile the family of solutions has a bifurcation point separating branches of growing shocks and of decaying shocks, depending on the density gradient and the initial velocity ratio between the shock and the local upstream (i.e., the ejecta just in front of the shock). We find a semi-analytic self-similar solution that separates these two branches, and show that it is unstable. 
We then use numerical simulations to map the family of solutions and use its results to find a full analytic description for the shock trajectory. Finally, we discuss aspects that can affect the observational signature of such shocks, particularly the altered density profile remaining behind the shock, and estimate the explosion energy required for decaying shocks to break out of a finite medium. 

We proceed as follows: in section \ref{sec: static} we revisit relevant existing solutions in a static medium. In section \ref{sec: expanding} we consider shock propagation in an expanding medium and, drawing on the static solution, show that there is a critical value of $\alpha$ below which all shocks decay, and above which two branches of solutions exist: one of growing shocks and another of decaying shocks, depending on the initial shock velocity. These regimes are separated by a self-similar solution, obtained semi-analytically in section \ref{sec: self-similar}, in which the ratio between the shock velocity and the velocity of the moving medium at the shock location is constant.  In section \ref{Density profile}, we derive the density profile remaining after the passage of the shock. In section \ref{simulations} we present numerical verification for our earlier analysis, and quantitative solutions for the shock propagation. Relying on results from the simulations, in section \ref{Analytic expression} we find full analytic expressions which describe the shock trajectory. We proceed by analysing the conditions for the shock to break out of the ejecta and produce a bright breakout emission in section \ref{sec: Breakout}. In section \ref{part: wind} we discuss the applicability of our solution to a shock driven by a fast wind, and in section \ref{part: summary} we summarize  our results.

\section{Shock propagation in a Static medium}
\label{sec: static}
Before proceeding with the characterization of spherical shock solutions in an expanding medium, let us examine first the existing solutions of a blast wave propagating in a static medium. In these solutions, the setup is as follows;  A spherical shock is driven into a spherically symmetric medium by a sudden release of energy in some small region at the center. This medium is assumed to be cold ideal gas with an adiabatic index $ \gamma $, and outside some small region at the center, the density has a power-law profile, $ \rho =Kr^{-\omega}$, where $r$ is the radius and $K$ and $\omega$ are constants. The asymptotic solutions for this set of problems are self-similar, taking the form $ \dot{R}=AR^\delta $, where $R$ is the shock radius, $A$ is a constant, and the value of $ \delta $ depends in the general case on the  power-law index $ \omega $ and the adiabatic index $\gamma$.

For a given value of $ \gamma $, the qualitative solution depends on the steepness of the density profile and can be divided into severl regimes.
For $ \omega< 3 $ the evolution of the shock wave follows the Sedov-Von Neumann-Taylor solution \citep{sedov1946propagation,von1947blast,Taylor1950}. The shock decelerates and stays in causal contact with the entire deposited energy and can be described by a self-similar solution of the first type. Energy conservation considerations, and dimensional analysis are sufficient to conclude that $\delta=-\frac{3-\omega}{2}$. 
 
For $\omega\ge \omega_g(\gamma)$ (where $ \omega_g\simeq3.13$ for $\gamma=\frac{4}{3}$, and $\omega_g\simeq3.26$ for $ \gamma=\frac{5}{3}$), there exists a second type self similar solution, originally derived by \cite{Waxman1993}. In this solution the shock accelerates, i.e., $ \delta > 0$, thereby losing causal contact with the bulk of the energy. The value of $\delta$ (from here on denoted $ 
\delta_{WS} $) should be found numerically. 
As we will see below, $\delta_{WS}=1$ is a critical value for the family of solutions of shocks propagating in expanding media. In a static media with   $\gamma=4/3$ this value is obtained for $\omega_c \simeq 8.22$ (for $ \gamma=\frac{5}{3}, \omega_c\simeq 7.69 $).

Since the adiabatic index $\gamma\ge1$, a gap exists between the two solutions. Inside this gap, additional second type self-similar solutions with $ \delta=0 $ exist, describing shock waves with a constant velocity \citep{Gruzinov2003,Kushnir2010}. 

\section{Shock propagation in a homologously expanding medium}
\label{sec: expanding}
Let us examine a spherical shock wave propagating in a freely expanding spherically symmetric medium with a power-law density profile of the form:
\begin{equation}
\rho_{ej} =K t^{-3}v^{-\alpha} ,
\label{density}
\end{equation}
where $K$ is some constant and $v=v(r,t)$ is the velocity of the ejecta. The unshocked ejecta is homologous, implying $v=r/t$ and that the internal energy of the medium is negligible in comparison with the kinetic energy, meaning the gas is cold. We consider only $\alpha > 5$, so that slow material carries more energy than faster material. We also assume that the gas is ideal.  
We consider adiabatic indices of $\gamma = 4/3$, which is relevant when the shock material is radiation dominated, and $\gamma = 5/3$ appropriate for cases in which the energy in the shocked matter is dominated by the gas. 
Our goal is to identify under which conditions shocks decay and under which conditions they grow. Once the initial conditions are forgotten, in addition to the adiabatic index $\gamma$ and the density power law $\alpha$, the shock evolution can only depend on the shock location, $R$, the time, $t$, the shock velocity $\dot{R}$, the local upstream velocity $v=\frac{R}{t}$, and the density normalization constant $K$. From consideration of dimensional analysis, $K$ cannot be a relevant parameter, and up to a scaling constant, the shock evolution must depend only on: $\gamma,\alpha$ and on the velocity ratio $\frac{\dot{R}}{v}=\frac{t\dot{R}}{R}$. We thus define: 
\begin{equation}\label{eq:eta_def}
    \eta \equiv \frac{\dot{R}}{v(R)}=\frac{d\log(R)}{d\log(t)},
\end{equation}
where in the second equality we invoked a ballistic ejecta with $v(R)=R/t$. 
We further define: 
\begin{equation}\label{eq:delta_def}
	\delta(\alpha,\eta) \equiv \frac{d\log (\dot{R})}{d\log{R}},
\end{equation}
In self-similar solutions (both in static and in expanding media) this index is a constant parameter which is determined in the solution and depends only on $\alpha$ (it is an eigenvalue of the shock equations). In the problem that we solve here, where the solutions are in general not self-similar, the velocity ratio $\eta$ and, hence, $\delta(\alpha,\eta)$ vary along the shock trajectory.  Nonetheless, $\delta$ varies slowly (over a dynamical time) and can serve to characterize the local expansion rate of the shock. This enables a comparison to the the self-similar solution in the static medium case. Moreover, if $\delta(\alpha,\eta)$ is known for every value of $\eta$ then the entire evolution of the shock can be found for every set of initial conditions. To elucidate this point, we express the evolution of $\eta$ with time as:
\begin{equation}\label{eq:deta_dt}
 \frac{d \log \eta}{d \log t} =\frac{d\log \dot{R}}{d \log R}\frac{d\log R}{d \log t}-\frac{d \log \frac{R}{t}}{d \log t}=  (\delta-1)\eta +1.
\end{equation}
It is now seen that if $\delta(\alpha,\eta)$ is known then for a given $\alpha$ and an initial value of $\eta$ at some time $t_0$, equation \ref{eq:deta_dt} can be integrated to yield $\eta(t)$. Then 
rearranging and integrating equation \ref{eq:eta_def} one obtains:
\begin{equation}
    R(t)=R(t_0)\exp \left[\int_{t_0}^t \frac{\eta(t')}{t'} dt' \right] ~.
\end{equation}
Thus, $\delta(\alpha,\eta)$ is an implicit solution of the shock evolution. We find it numerically for various values of $\alpha$ and $\eta$ in \ref{simulations}, and find a full analytic expression for it in \ref{Analytic expression}.

Since there are two characteristic velocities in the problem we do not expect to find, in general, self-similar solutions. However, there are two cases for which self-similar solutions may exist. The first one is a family of exact solutions derived for a unique configuration in which  the velocity ratio $\eta$ is constant along the entire shock trajectory. As we shall show later, there are values of $\alpha$ for which such solutions exist, though they are unstable. The second one, is approximate solutions 
 obtained in the limit where the medium velocity $v$ is negligible ($\eta \gg 1$). These solutions approach the static medium solution  $\delta(\alpha,\eta \rightarrow \infty)= \delta_{WS}(\omega=\alpha)$.  
 The convergence of these solutions to the static medium case stems from the fact that when $\eta \to \infty$ the motion of the ejecta over a dynamical time of the shock is negligible. The fact that the medium is moving affects the value of $\eta$ since neglecting the medium motion over a dynamical time implies that we can approximate $v(r=R)\propto R$. Using this approximation we obtain a simple formula for the evolution of $\eta$ in this limit:
\begin{equation}
\eta\gg 1:     \eta=\frac{\dot{R}}{v}\propto\frac{\dot{R}}{R}\propto R^{\delta-1}.
\end{equation} 
As we show next, solutions are separated into two types. One type corresponds to growing shocks, for which $\eta$ increases monotonically with time; as $\eta \to \infty$  these solutions approach asymptotically the Waxman-Shvarts solutions in a static medium. The second type consists of decaying shocks for which $\eta$ declines with time. We find that in all decaying shock solutions the shock velocity ultimately approaches the local velocity of the moving medium, viz., $\dot{R}\to v(R)$ ($\eta\to1$), at which point the shock dies out. Equivalently, we may classify a shock as decaying if there exists a mass shell that the shock will never cross. This condition is different than in the static case, in which a decaying shock decelerates, and its velocity approaches zero in the lab frame.

It is worth noting that as viewed in the lab frame, both decaying and growing shocks accelerate. Also in the local upstream frame, the distinction between accelerating and decelerating shocks does not coincide with the distinction between growing and decaying shocks.  For example, a shock with constant $\eta=\eta_c>1$ will accelerate in the upstream frame; $\frac{d}{dt}(\dot{R}-R/t)=\frac{d}{dt}(\dot{R}(1-1/\eta_c))>0$. Similarly, a decaying shock with $\eta$  slightly smaller than $\eta_c$ will also accelerate in the local upstream frame.

The evolution of the shock, growing or decaying, is determined by the sign of $\dot{\eta}$. Examining equation \ref{eq:deta_dt} at the limit of $\eta \to \infty$, the sign of $\dot{\eta}$ depends solely on the sign of $\delta-1$. In this limit $\delta \simeq \delta_{WS}$, therefore if $\delta_{WS} >  1$, meaning $\alpha > \omega_c$, and  $\eta \to \infty$ then $\dot{\eta}>0$ and shocks grow. If $\delta_{WS}<1$,  which corresponds to $\alpha <\omega_c$, then for large values of $\eta$ the shocks decay. 
In each of these regimes if the behaviour exhibited for $\eta \to \infty$ does not persist for all values of $\eta$ (i.e., $\dot{\eta}$ changes sign), then there exists some value of $\eta$ for which $\dot{\eta}=0$, meaning there is a solution with a constant $\eta=\eta_c$. In the next section we show that when $\eta=\eta_c$ the hydrodynamic equations admit a self similar solution. We farther show that such a solution exists only for $\alpha\ge \omega_c$, hence, for $\alpha<\omega_c$ shocks always decay, that is, for every set of initial conditions, $\eta$ declines along the shock trajectory until the shock dies out.

For $\alpha>\omega_c$ we find that there is a solution with $\eta=\eta_c$. Based on our discussion above we can see that this solution is unstable, since $\dot{\eta}>0$ for $\eta>\eta_c$. In our numerical simulations we find that over the entire parameter space examined (values of $\alpha$ between 6 and 14) $\dot{\eta}<0$ for $\eta<\eta_c$. Thus we conclude that for\footnote{In our self similar solution we verify that $\eta_c(\alpha)>1$ at least for every $\alpha \le 20$.} $\eta_c>1$ the self-similar solutions with $\eta=\eta_c$ are unstable, and bifurcates into two branches: growing shocks ($\eta>\eta_c$) and decaying shocks ($\eta < \eta_c$).

To summarize, depending on the value of $\alpha$, the shock behaviour can be:
\begin{center}
	\begin{tabular}{ |c| c | c| } 
		\hline
		$ \alpha<\omega_c $ &\multicolumn{2}{c|}{$\alpha\ge\omega_c$} \\
		\hline
		 \multirow{3}{6em}{shock decays } & $ \eta<\eta_c $ & shock decays \\ 
		 & $ \eta = \eta_c $ & self-similar solution \\
		 & $ \eta >\eta_c $ & shock grows \\
		\hline
	\end{tabular}
\end{center}

\subsection{Self-similar solutions for the limiting case}
\label{sec: self-similar}

We seek a self-similar solution where for a given value of $\alpha$ the ratio between the shock velocity and the local velocity of the medium is constant along the trajectory of the shock, $\eta = \eta_c(\alpha)$. Before proceeding with the analysis, let us emphasise some basic properties of the desired solution. 

First, as discussed above, such a solution exists only for $ \alpha>\omega_c $ and it is unstable. Second from equation \ref{eq:deta_dt} we see that for $\eta=\eta_c$ that is constant, $\delta=\frac{\eta_c-1}{\eta_c}=const$. Therefore,
\begin{equation}
\dot{R}=AR^\delta.
\label{R_sim}
\end{equation}
Note that $ \delta $ is always smaller than 1, and monotonically increases with $\eta_c$,  approaching 1 for  $ \eta_c \to \infty $. When $\eta_c\to \infty$, the self similar solution coincides with the static case solution, in which $\delta_{WS}=1$ is obtained for $\alpha=\omega_c$, therefore, $ \eta_c \to \infty $  as $\alpha\searrow\omega_c$, where the tilted arrow denotes that the limit is from above. 
It is important to realize that in case of an expanding medium, the local density gradient traversed by the shock is not a power-law with index $\alpha$, but rather an effective density gradient which depends on the original power law as well as the shock velocity. By taking the logarithm of Eq. \ref{density} and then differentiating it with respect to $\log(R)$, we obtain the effective local density power-law seen by the shock:
\begin{equation} 	\label{omega_eff}
	\omega_{eff} \equiv -\frac{d\log (\rho)}{d\log{R}}=\alpha -(\alpha-3)\frac{R}{\dot{R}t}=  \alpha - \frac{\alpha-3}{\eta}.
\end{equation}
Note that since $\alpha>3$, the shock samples a weaker density profile than the one measured in the lab frame; $ \omega_{eff}<\alpha $. Plugging $\eta_c$ into equation \ref{omega_eff}, we find that the effective density profile seen by the shock is given by $\omega_{eff}=3-3\delta+\alpha\delta=const $. 

Expressing the ejecta density at the shock location in terms of the constants defined above we have
\begin{equation}
\rho_{ej}(r=R) =B R^{-\omega_{eff}},
\end{equation}
where $ B =  K ((1-\delta)A)^{3-\alpha} $. 
We now proceed by defining a similarity variable  $ \xi = \frac{r}{R} $. Writing the flow fields behind the shock in  terms of the similarity variable and the shock velocity $ \dot{R} $, the shock structure is not time dependent, and has a single length scale $ R $:
\begin{equation}
\begin{split}
v(r,t) &= \dot{R}\xi U(\xi),\\ c_s(r,t) &= \dot{R}\xi C(\xi),\\ \rho (r,t)  &= BR^{-\omega_{eff}}H(\xi),
\end{split}
\label{flow_params}
\end{equation} 
here $c_s$ is the speed of sound.

For the self-similar system, Eqs. \ref{R_sim} and \ref{flow_params}, the hydrodynamic equations for an adiabatic flow,
\begin{align}
\left(\partial_t+v \partial_r\right)\ln\rho +\frac{1}{r^2}\partial_r\left(r^2v\right) &=0,\\
\left(\partial_t+v \partial_r\right)v+\frac{1}{\rho}\partial_r\left(\frac{\rho c_s^2}{\gamma}\right) &=0,\\
\left(\partial_t+v \partial_r\right)\frac{c_s^2}{\rho^{\gamma-1}} &=0,\\
\end{align}
reduce to a differential equation and a quadrature:
\begin{align}
\frac{d U}{d C} &= \frac{\Delta_1(U,C)}{\Delta_2(U,C)} \label{dUdC},\\
\frac{d U}{d \ln (\xi)} = \frac{\Delta_1}{\Delta}  \quad &\textrm{or} \quad \frac{d C}{d \ln (\xi)} = \frac{\Delta_2}{\Delta},
\label{quad}
\end{align}
where:
\begin{align}
\Delta &= C^2 - (1-U)^2, \\ \label{delta}
\Delta_1 &= U(1-U)(1-U-\delta) -C^2\left(3U+\frac{2\delta - \omega_{eff}}{\gamma}\right),\\ 
\begin{split}
\Delta_2 &= C \Bigl[(1-U)(1-U-\delta) \\
& \quad - \frac{\delta-1}{2} (2-2U+\delta) U +  \left(\frac{2\delta+(\gamma-1)\omega_{eff}}{2\gamma(1-U)}-1\right) C^2\Bigr],
\end{split}  \label{delta2}
\end{align}
and H is given implicitly by
\begin{equation}
H^{\gamma-1+\lambda}\left(\xi C\right)^{-2}\left(1-U\right)^{\lambda}\xi^{3\lambda}=\text{const,}
\label{H implicit}
\end{equation}
with $ \lambda=\frac{2\delta+\left(\gamma-1\right)\omega_{eff}}{3-\omega_{eff}}$. 

To solve these equations, boundary conditions at the shock front $ \xi=1 $ are needed. The boundary conditions $U(1),C(1),H(1)$ are obtained from the Rankine-Hugoniot Jump conditions (we use subscript '1'['2'] for upstream [downstream] quantities),

\begin{eqnarray}
\rho_1\left(\dot{R}-v_1\right) = \rho_2\left(\dot{R}-v_2\right) \label{RH1}, \\ 
\rho_1 v_1\left(\dot{R}-v_1\right)-\frac{\rho_1 c_{s,1}^2}{\gamma} = \rho_2 v_2\left(\dot{R}-v_2\right)-\frac{\rho_2 c_{s,2}^2}{\gamma} \label{RH2}, \\
\rho_{e_1} \left(\dot{R} -v_1\right)\left(\frac{v_1^{2}}{2}+\frac{c_{s,1}^{2}}{\gamma-1}\right) - \frac{\rho_{1}c_{s,1}^{2}\dot{R}}{\gamma} \\ \nonumber
= \rho_2 \left(\dot{R}-v_2\right)\left(\frac{v_2^{2}}{2}+\frac{c_{s,2}^{2}}{\gamma-1}\right)-\frac{\rho_2 c_{s,2}^{2}\dot{R}}{\gamma},
\label{RH3}
\end{eqnarray}
and read: 
\begin{align}
U(1) &= \frac{2+(\gamma-1)(1-\delta)}{\gamma+1} \label{boundU},\\
C(1) &= \frac{\sqrt{2\gamma(\gamma-1)}}{\gamma+1}\delta \label{boundC},\\
H(1) &= \frac{\gamma+1}{\gamma-1}.
\end{align}

For solutions to exist and be single-valued, they must pass the sonic point $ \Delta = 0 $ at the singularity $ \Delta_1=\Delta_2=0 $. This fixes $ \delta $ for every value of $ \alpha $. This point is given implicitly by the following equations, and it can be verified that $ \Delta_1,\Delta_2 $ vanish identically.
\begin{align}
C &= 1-U, \\
U(1-U-\delta) &= (1-U)\left(3U+\frac{2\delta-\omega_{eff}}{\gamma}\right).
\end{align}

\begin{figure}
	\center
	\includegraphics[width=0.5\textwidth]{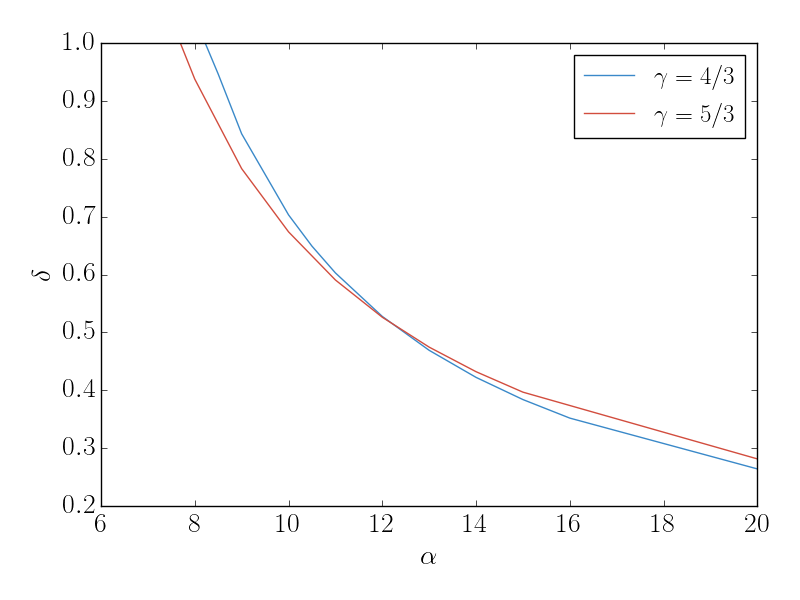}
	\caption{ The dependence of $\delta$ on $\alpha$ derived from the self-similar solutions. 
	When $\alpha$ approaches $\omega_c$, $\delta$ approaches 1.}
	\label{fig:delta(alpha)}
\end{figure}

\begin{figure}
	\center
	\includegraphics[width=0.5\textwidth]{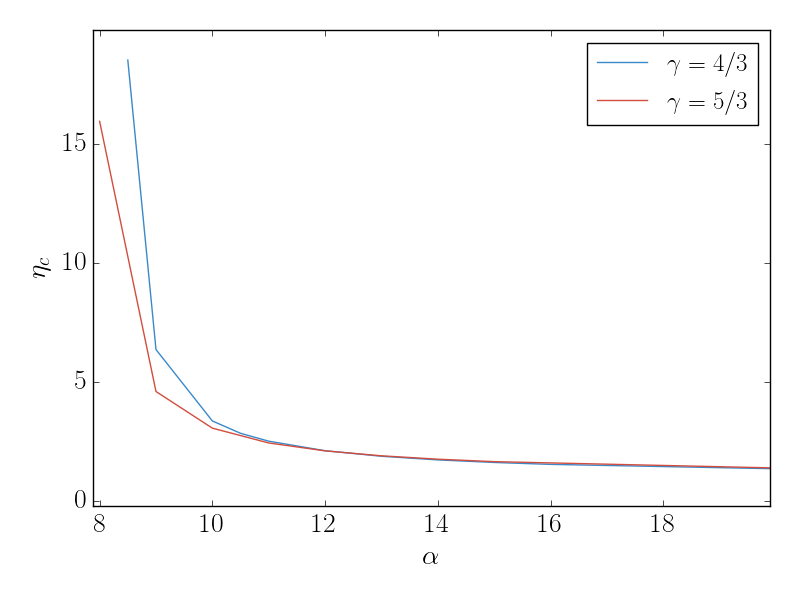}
	\caption{ $\eta_c$ as a function of $\alpha$, as obtained from the self-similar solutions.}
	\label{fig:eta(alpha)}
\end{figure}

To find the eigenvalue $\delta$ for a given setup (i.e., given values of $\alpha$ and $\gamma$), we numerically integrate Eq. (\ref{dUdC}) and one of the equations in (\ref{quad}) starting from the boundary conditions at the shock front (Eqs. (\ref{boundU}),(\ref{boundC})), and search for the value of $\delta$ for which the solution passes through the sonic point. The dependence of $\delta$ on $\alpha$ obtained from the family of self-similar solutions (for which $\eta=\eta_c$) are shown in figure \ref{fig:delta(alpha)}. The value of $\eta_c$ as a function of $\alpha$ is shown figure \ref{fig:eta(alpha)}. It shows that, as expected, $\eta_c \to \infty$ when $\omega \searrow \omega_c$ and on the other end $\eta_c \to 1$ for large values of $\alpha$. The drop in $\eta_c$ is very sharp where already for $\alpha=10$ we obtain $\eta_c \approx 3$ for both adiabatic indices that we consider.

\subsection{The far downstream density profile}
\label{Density profile}
The passage of the shock alters the distribution of the pre-shocked outflow velocity. The new distribution may have a direct impact on the observed signature in cases where the observed photons diffuse through the shocked medium, and in cases where the shocked outflow itself generates radiation by interaction with some external medium.
Once fluid elements cross the shock they start rarefying, converting their internal energy to bulk motion. Immediately behind the shock the internal energy is at most comparable to the bulk energy and therefore after the entire internal energy is converted to bulk motion the velocity does not change by much.  Consequently, the velocity in the far downstream, $v_f$, is roughly the same as the velocity in the immediate downstream, $v_f
\simeq \dot{R}$. 

As a result, and given that the shock enhances the density by a constant factor, the density at the shock location scales with the shock velocity in the same manner as the density far behind the shock scales with the terminal velocity, viz., $\frac{d \log \rho }{d \log \dot{R}} \approx \frac{d \log \rho_f}{d \log v_f}$. 
In the static limit ($\eta \to \infty$), the pre-shock density profile is $\rho \propto r^{-\alpha}$, therefore,   $\rho(R)\propto R^{-\alpha}$. Since $\delta= \delta_{WS}(\omega=\alpha)$, we find $\rho(\dot {R})\propto R^{-\frac{\alpha}{\delta_{WS}}}$, and since for a given fluid element $v_f \approx v \propto R $ we obtain $\rho_f\propto v_f^{-\frac{\alpha}{\delta_{WS}}}$.
For the general case of expanding medium, $\rho\propto v^{-\alpha}$, and $v=\frac{\dot{R}}{\eta}$. Therefore, $\rho(\dot{R})\propto\eta^\alpha \dot{R}^{-\alpha}$ and $\rho_f \propto \eta^\alpha v_f^{-\alpha}$, where $\eta$ is evaluated when the fluid element with $v_f$ passes the shock: 
\begin{equation}
\frac{d \log \rho_f }{d \log v_f}=\frac{d\log \rho }{d\log \dot{R}}=-\alpha + \frac{\alpha}{\delta} \frac{d \log \eta }{d \log R}.
\label{eq:rho_f_exp}
\end{equation}
 In the self-similar case, $\eta=\eta_c=const$, and the density profile far behind the shock is the same as the original density profile, $\rho_f(\eta=\eta_c) \propto v_f^{-\alpha}$.  In accelerating shocks, the second term on the R.H.S of Eq. (\ref{eq:rho_f_exp}) is positive, hence the final density profile will be shallower than the initial profile, approaching the static limit, $\rho_f(\eta \to \infty) \propto v_f^{-\alpha/\delta_{WS}}$. In decaying shocks, the second term is negative, and becomes more negative as  $\eta\to1$ and $\delta \to 0$, so that 
 the profile is steeper than the initial density profile. Note that in this limit, since the shock is converging to a fixed location (in a Lagrangian sense) within the ejecta, the steep density profile zone is confined to a small section of the ejecta which was crossed by the shock prior to its complete decay. A full analytic expression for equation \ref{eq:rho_f_exp} is presented in section \ref{Analytic expression} in equation \ref{downstream_density}. 
 
\begin{figure}
	   \center
	   \includegraphics[width=0.5\textwidth]{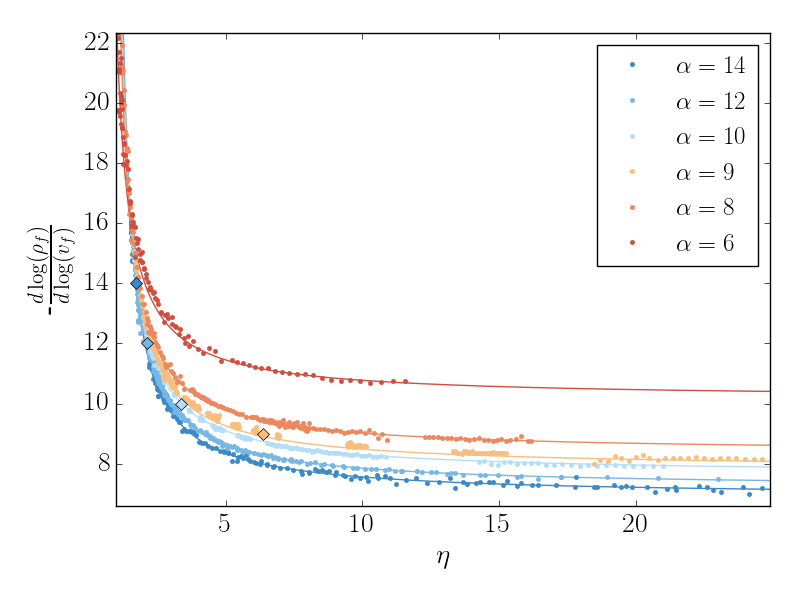}
	\caption{Logarithmic derivative of the far down stream density as a function of $\eta$ (evaluated when the fluid element with $v_f$ has crossed the shock). The values corresponding to the self-similar solutions, $-\frac{d\log \rho_f}{d \log v_f} = \alpha$, are marked by diamonds, and the solid lines mark the analytic expression in eq. \ref{downstream_density}}.
	\label{fig:dlogrho_f dlogv_f}
\end{figure}

In Fig. \ref{fig:dlogrho_f dlogv_f} the dependence of $\frac{d\log \rho_f}{d \log v_f}$ on $\eta$ (found numerically) is plotted for different values of $\alpha$. For growing shocks, the density profile quickly approaches that of the static case, and the resulting profiles are similar and not strongly affected by the initial density profile. For example, when $\alpha$ changes from 6 to 14 the value of $\frac{\alpha}{\delta_{WS}}$ changes roughly from 7 to 10.

\subsection{Numerical simulations}
\label{simulations}
We carried out numerical simulations of spherical shocks driven into expanding gas. The simulations start with 'piston' which is used to drive a shock into the expanding ejecta. Denoting the code units by $\rho_0,v_0,r_0$, the unit pressure and unit time can be expressed as $t_0=\frac{r_0}{v_0},P_0=\rho_0 v_0^2$. 
The initial conditions are given by:
\begin{equation}
\begin{aligned}
\rho_i & = 
\rho_0 \begin{cases}
1, r_0< r\le 2r_0\\
\left(\frac{r}{r_0}\right)^{-\alpha}, r>2r_0
\end{cases}\\
P_i &= f_0\cdot\left (\frac{\rho_i}{\rho_0}\right) ^{\gamma} P_0\\
v_i &= \frac{r}{t'} 
\end{aligned}
\end{equation}
where $f_0 = 10^{-4}$ for $\gamma = \frac{4}{3}$, $f_0 = 10^{-6}$ for $\gamma =\frac{5}{3}$, and $t'$ is a constant measured in time units. 
The "piston" is created by pushing dense ($\rho(r=1) = 10^{6}\rho_0$) cold  matter through the $r=r_0$ boundary. The piston velocity determines the shock velocity, and is varied between simulations, in order to scan a range of $\eta$  values. 
An ideal gas equation of state is used, and we carry out a set of simulations with $\gamma=\frac{4}{3}$ and a set of simulations with $\gamma=\frac{5}{3}$. The shock is tracked from the radius at which it loses causal contact with the piston, i.e, when the first grid cell that obeys $c + v < \dot{R}$ is in the ejecta, and till it's radius increases by 1-2 orders of magnitude. We verified that in this range, if the piston is stopped, the evolution of the shock is not affected. 

Simulations were carried out using the hydrodynamic module of PLUTO, version 4.2 \citep{Mignone_2007}. The simulations are 1D spherically symmetric, and exploit parabolic reconstruction and a third order Runge-Kutta time stepping. A uniform grid was used, ranging between $r=r_0$ and $r=(100-1000)r_0$, with a cell density ranging between 16 and 50 cells per unit length of $r_0$. Convergence was verified for a simulation with a low grid density (16 cells per unit of $r_0$) by running four simulations with the same initial conditions; a simulation with double the grid cells, one with half of the grid cells, and one with a quarter the grid cells. The first two of these agreed with the simulation used on all parameters used in the figures. 

The shock location and time are collected from each simulation.  Then the ejecta velocity at the shock location is calculated, and the shock velocity is found by numerical differentiation. The rest of the variables are derived, and the simulations are grouped according to the value of $\alpha$.  Note that for every value of $\alpha$ there is an overlap in $\eta$ values between different simulations. The agreement between different simulations  when the values of $\eta$ overlap is another indication that the shocks in the simulations have forgotten the initial conditions. 


\begin{figure}
	   \center
	   \includegraphics[width=0.5\textwidth]{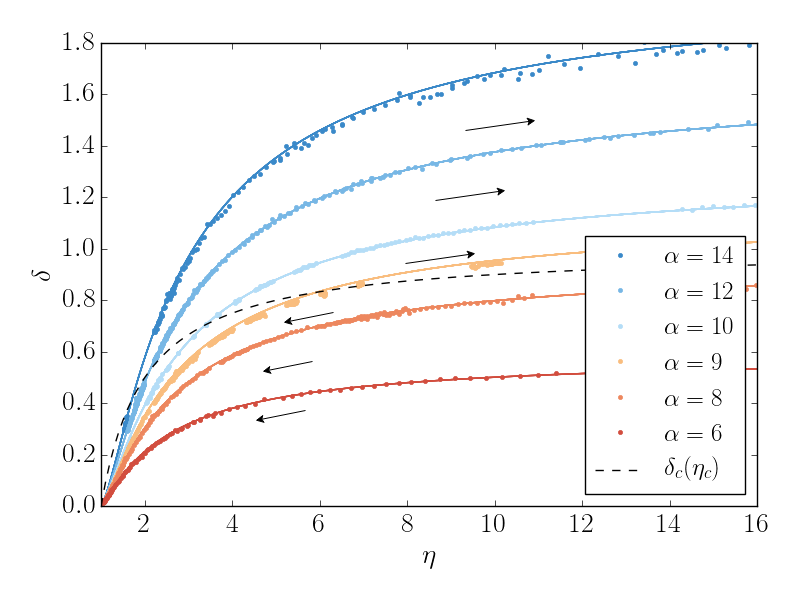}
	\caption{ Dependence of $ \delta $ on $ \eta $ for $\gamma$=$\frac{4}{3}$ and different values of $ \alpha $. The values from the numerical simulations marked by dots, and the solid line marks the analytic expression presented in eq. \ref{eq: delta}. The black dashed line marks the values of the self-similar solutions. This line crosses $ \alpha=9,10,12,14>\omega_c $ dividing each curve into two regimes, decaying shocks below, and growing shocks above. Shocks with $\alpha=6,8<\omega_c$ are always below the line and thus are always decaying.}
	\label{fig:delta(eta)}
\end{figure}

\begin{figure}
	   \center
	   \includegraphics[width=0.5\textwidth]{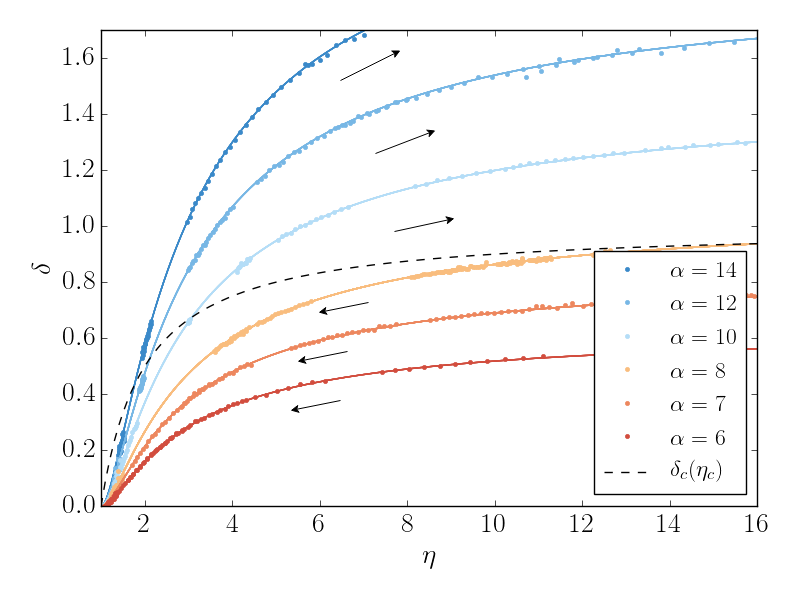}
	\caption{Same as figure \ref{fig:delta(eta)} for $\gamma$=$\frac{5}{3}$. Here $\omega_c \simeq 7.69$. Thus curves for  $ \alpha=8,10,12,14>\omega_c$ cross the black lack dashed line that marks the self-similar solutions and divides each curve into two regimes, decaying shocks below, and growing shocks above. Shocks with $\alpha=6,7<\omega_c$ are always below the line and thus are always decaying }
	\label{fig:delta(eta)53}
\end{figure}

\begin{figure}
	\center
	\includegraphics[width=0.5\textwidth]{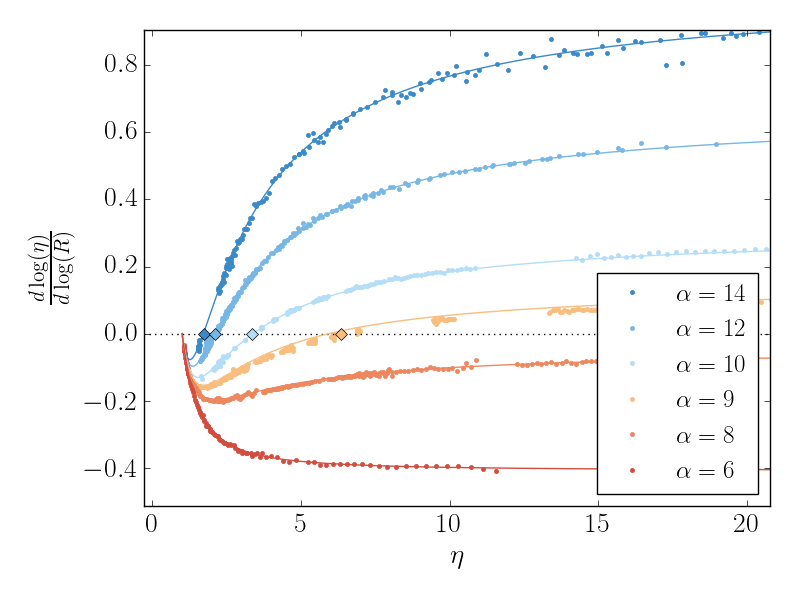}
	\caption{ Dependence of $ \frac{d\log\eta}{d\log R} $ on $ \eta $ for different $ \alpha $ plotted in colored dots. note that for $\alpha=6,8$, $\frac{d\log\eta}{d\log R}$ is always negative and the shock decays while for $ \alpha=9,10,12,14 $, there is a region in which the shock becomes stronger and grow and a region where the shock becomes weaker and decay. The black-framed markers mark the semi-analytical value of $\eta_c$, and the solid lines show the analytic expression derived in section \ref{Analytic expression}.}
	\label{fig:etadot(eta)}
\end{figure}

Fig \ref{fig:delta(eta)} (\ref{fig:delta(eta)53}) shows the dependence of $ \delta $ on $ \eta $ and $ \alpha $ for $\gamma = 4/3$ ($\gamma = 5/3$). The black dashed line marks the self similar solution, and divides the parameter plane into two regimes:  The domain under the black line corresponds to shocks that decay as $\eta$ drops with time, which are described by trajectories moving downwards and left with time (towards smaller $ \delta$ and $\eta $). In the region above the dashed line, shocks move upwards and to the right with time as they strengthen. This can also be seen by looking at the derivative of $ \eta $ with respect to the shock trajectory (Fig \ref{fig:etadot(eta)}).

\subsection{Full analytic expression for the shock propagation}
\label{Analytic expression}
\begin{figure}
	   \center
	   \includegraphics[width=0.5\textwidth]{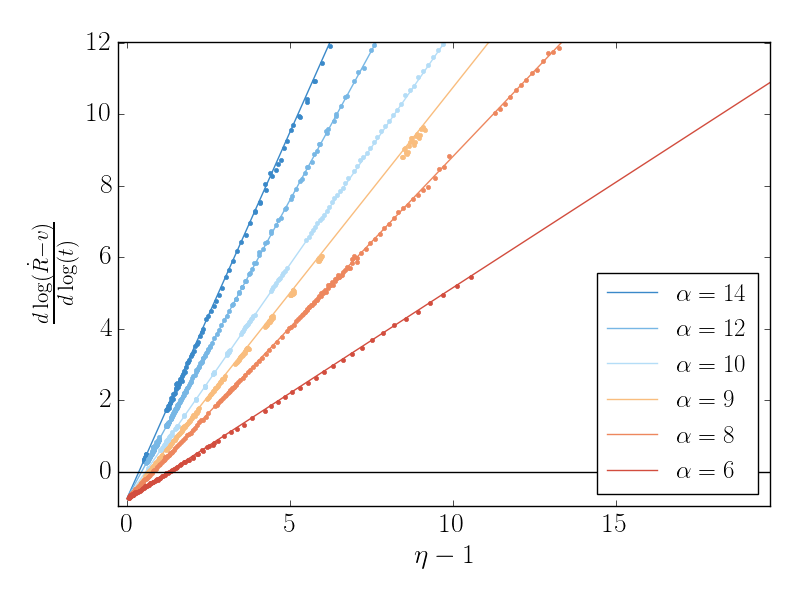}
	\caption{ Dependence of $ \frac{d \log \left(\dot{R}-v\right)}{d \log t} $ on $ \eta - 1$ for $\gamma$=$\frac{4}{3}$ and different values of $ \alpha $. This figure shows that the relation between these two parameters is linear and that lines for different values of $\alpha$ have the same intercept.}
	\label{fig:linear}
\end{figure}

Examining figure \ref{fig:linear}, it seems that $\frac{d\log(\dot{R}-v)}{d \log t}$ is linear in $(\eta-1)$, and that the graphs for different values of $\alpha$ share the same intercept. While we do not have an explanation for the linear behavior seen in the graph, the common intercept which exists for different values of $\alpha$ is not surprising. As $\eta\to 1$, the shock location doesn't change much in the Lagrangian sense. Hence, it is no longer affected by the density profile ahead of it. All the while, the mass causally connected to the shock decreases, and and structure it had no longer affects the shock.

Under the conjecture that this expression is indeed linear,
\begin{equation}
    \frac{d\log(\dot{R}-v)}{d \log t}=a (\eta-1) - b,
    \label{dlogv_rel}
\end{equation}
we proceed by finding $\delta(\eta), \eta(t),R(t)$. The main results of this derivation are summarized here, for the complete details, please see appendix \ref{Appendix A}. 
We start by simplifying equation \ref{dlogv_rel} to the following form:

\begin{equation}
    \frac{d\log(\eta-1)}{d \log t}=(a-1)(\eta-1) - b.
    \label{eq:dlogeta-1}
\end{equation}
Before solving this equation to find the explicit dependence of $\eta$ on time, we recall the equation for $\dot{\eta}$ (Eq. \ref{eq:deta_dt}), and comparing the two, we find:
\begin{equation}
\delta =\frac{\left(\eta-1\right)\left(\delta_{WS}\left(\eta-1\right)-b+1\right)}{\eta^{2}}
\label{eq: delta}
\end{equation}
Where $a=\delta_{WS}$ was determined by examining the limit $\eta \to \infty, \delta \to \delta_{WS}$.
Returning to Eq. \ref{eq:dlogeta-1}, for $\delta_{WS}>1$, when $\eta=\eta_c$ the left hand side of the equation must vanish. Hence, we find $b=(\eta_c-1)(\delta_{WS}-1)$. Plugging in the values from the semi-analytical solution, we find that $b\simeq 0.75$ for $\gamma = 4/3$, and $b\simeq 1$ for $\gamma =5/3$.  
Note that $b$ is positive, which means that for every $\alpha$ there exists a regime of $\eta$ values for which $\dot{\eta}<0$, and shocks decay.  

When the left hand side does not vanish, by solving this ODE we find:
\begin{equation}
    \eta=1+\frac{b\left(\eta_{i}-1\right)\left(\frac{t}{t_{i}}\right)^{-b}}
    {b-\sigma(t)}
\end{equation}
Where $t_i$ is set as the time at which $\eta=\eta_i$, and $\sigma(t)=\left(\delta_{WS}-1\right)\left(\eta_{i}-1\right)\left(1-\left(\frac{t}{t_{i}}\right)^{-b}\right)$
Plugging in $\eta = \frac{\dot{R}t}{R}$, we may integrate to find:
\begin{equation}
\frac{R}{R_{i}}=\begin{cases}
\frac{t}{t_{i}}\left(\frac{b}{b+\left(\frac{t}{t_{i}}\right)^{-b}\sigma(t)}\right)^{\frac{1}{\delta_{WS}-1}}, &\delta_{WS}\neq1\\
\frac{t}{t_{i}}\exp\left(\frac{\left(1-\left(\frac{t}{t_{i}}\right)^{-b}\right)(\eta_i-1)}{b}\right), &\delta_{WS}=1
\end{cases} 
\label{R(t)}
\end{equation}
Where $R_i=R(t=t_i)$. 
Examining $\dot{R}(t)$, 
\begin{equation}
\begin{split}
\dot{R} &= R_i \left(1+\frac{b h_{i}\left(\frac{t}{t_{i}}\right)^{-b}}{b-\sigma(t)}\right) \times \\
& \times\begin{cases}
\left(\frac{b}{b+\left(\frac{t}{t_{i}}\right)^{-b}\sigma(t)}\right)^{\frac{1}{\delta_{WS}-1}}, &\delta_{WS}\neq1\\
\exp\left(\frac{\left(1-\left(\frac{t}{t_{i}}\right)^{-b}\right)(\eta_i-1)}{b}\right), &\delta_{WS}=1
\end{cases} 
\end{split}
\end{equation}
we can see that for $\delta_{WS}>1$, the shock grows if the solution passes through the singularity $b-\sigma(t)=0$, meaning $\frac{b}{\delta_{WS}-1}+1<\eta_i$. In this case, the shock will diverge in finite time: 
\begin{equation}
t(R\to \infty) =t_i \left(\frac{\eta_{i}-1}{\eta_{i}-\eta_c}\right)^{\frac{1}{b}}
\end{equation}
Where we used $b=(\delta_{WS}-1)(\eta_c-1)$.

For dissipating shocks, regardless of whether $\delta_{WS}$ is larger of smaller than unity, taking the limit $t\to \infty$, we find that velocity approached by the shock is:
\begin{equation}
\lim_{t\to \infty}\dot{R}
=\frac{R_{i}}{t_i} \begin{cases}
\left(\frac{b}{b-\left(\delta_{\text{WS}}-1\right)(\eta_i-1)}\right)^{\frac{1}{\delta_{WS}-1}}, & \delta_{WS}\neq1\\
\exp\left(\frac{\eta_{i}-1}{b}\right), & \delta_{WS}=1
\end{cases}
\end{equation}Relying on our previous results, we can also find:
\begin{equation}
\frac{d\log \eta}{d \log R} = \frac{\eta-1}{\eta^{2}}\left(\left(\delta_{WS}-1\right)\left(\eta-1\right)-b\right),
\label{deta drho expr}
\end{equation}
which is plotted in figure \ref{fig:etadot(eta)} alongside the solution from the simulations. 
This expression can be used to derive the far downstream density profile described in eq. \ref{eq:rho_f_exp}. 
\begin{equation}
\frac{d \log \rho_f }{d \log v_f}=\frac{d\log \rho }{d\log \dot{R}}= -  \alpha\frac{\eta}{1-b+\left(\eta-1\right)\delta_{\text{WS}}},
\label{downstream_density}
\end{equation}which is plotted in figure \ref{fig:dlogrho_f dlogv_f}. 

For convenience, we summarize the values of $a =\delta_{WS}$  and of b in the following table:
\begin{center}
	\begin{tabular}{ |c| c | c| } 
		\hline
		&\multicolumn{2}{c|}{a=$\delta_{WS}$} \\
		\hline
        $\alpha$ & $\gamma = 4/3$ & $\gamma = 5/3$ \\
        \hline
		6 & 0.6 & 0.6\\
        7& 0.8 & 0.9\\
        8 & 0.9 & 1\\
        9 & 1.2 & 1.3\\
        10 & 1.3 & 1.5\\
        12& 1.6 & 1.9\\
        14& 2.1 & 2.3\\
		\hline
        &\multicolumn{2}{c|}{$b$} \\
        \hline
        & 0.75 & 1\\
        \hline
	\end{tabular}
\end{center}

\subsection{Shock breakout}

\label{sec: Breakout}
When the shock breaks out, the radiation trapped within the shock is released to the observer, emitting a bright flare. This happens when the optical depth to the observer drops below $\tau \approx \frac{c}{v}$. It is therefore interesting to understand in which systems a shock will break out and in which it will die out before reaching the breakout radius.

For a growing shock, the answer is simple, as it will always break out eventually. For decaying shocks the answer is far more complex since some will die out at a radius for which the optical depth to infinity is still larger than $\frac{c}{v}$, while others will break out. For a given decaying shock, the condition for breakout depends on the optical depth of the medium at the mortality radius, and, therefore, it has to be calculated for each system separately. However, since every outflow has a maximal expansion velocity, $v_{\max}$, we can address the question of what are the conditions required for a decaying shock to reach $v_{\max}$, in which case it is guaranteed that it will break out successfully. It turns out that for a given $\alpha$ the criterion for reaching $v_{\max}$ depends on two dimensionless parameters, the ratio between the explosion energy and the outflow energy, $E_{exp}/E_{ej}$, and the ratio between the maximal ejecta velocity and the velocity at the onset of the steep density gradient tail (i.e., the velocity that carries most of the ejecta energy), $v_{\max}/v_0$.      

\begin{figure}
	\center
	\includegraphics[width=0.5\textwidth]{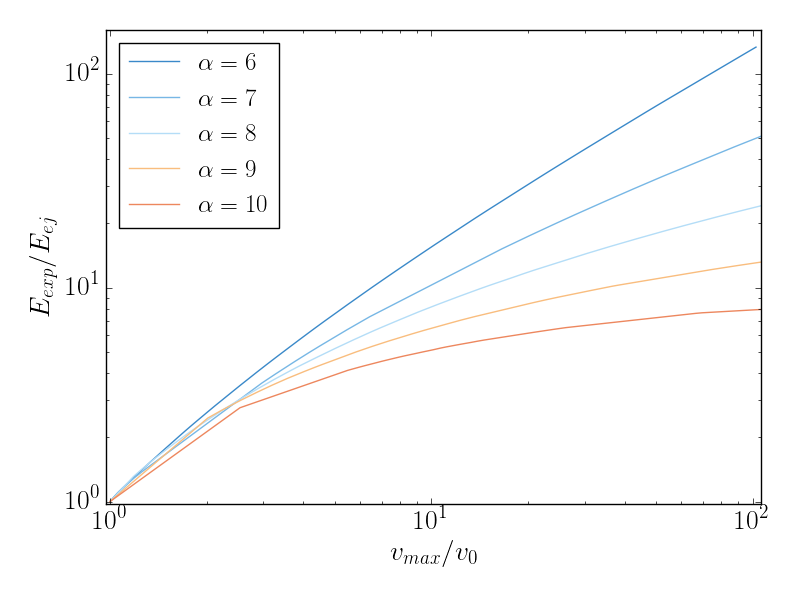}
	\caption{The minimum energy ratio needed for a shock breakout from an expanding medium of a finite extent. The velocity of the outermost shell is $v_{max}$, the velocity of the ejecta at the onset of the steep density gradient tail is $v_0$, $E_{exp}$  is the explosion energy, and $E_{ej}$ is the
       ejecta energy. If the shock reaches an optical depth of $\tau = \frac{c}{\dot{R}}$ before reaching the edge of the ejecta, it will breakout earlier.}
	\label{fig:E vs v}
\end{figure}

Figure \ref{fig:E vs v} shows the minimal energy ratio needed for a shock to reach $v_{\max}$ and break out of the ejecta. The curves were calculated as follows. The shock which is released at the center of the ejecta starts decaying as it traverses the shallow part of the ejecta and its velocity at the onset of the steep density gradient zone can be approximated by $\sqrt{2E_{exp}/m_{ej}}$, where we use the facts that in the shallow region the energy that is in causal contact with the shock is conserved and that the contribution of the steep part to the ejecta mass is negligible. The total ejecta energy can be approximated by $E_{ej}\approx m_{ej} v_0^2/2$. Therefore,  the shock starts its accelerating phase, where the unshocked ejecta velocity is $v_0$, with  $\eta \approx \sqrt{\frac{E_{exp}}{E_{ej}} }$. The value of $\frac{v_{max}}{v_0}$ is then found numerically by integrating $\frac{dv}{d\eta}$, starting at $v_0$, and finding $v_{max}$ where $\eta \to 1$. When examining Fig. \ref{fig:E vs v} it is seen, first, that as expected for $\alpha>\omega_c$, where shocks with $\eta>\eta_c$ are not decaying, the value of $E_{exp}/E_{ej}$ that guarantees breakout converges to a finite value, $\eta_c(\alpha)^2$, when $v_{\max}/v_0 \to \infty$. Second, in most systems, the ratio $v_{\max}/v_0$ is not very large. For example in a SN explosion of an extended star such as a red supergiant $v_{\max}/v_0 \approx 5$ while in SN explosion of a compact Wolf-Rayet star  $v_{\max}/v_0 \approx 20$ \citep[e.g.][]{Nakar2010a}. This implies that an energy ratio of $E_{exp}/E_{ej} \approx 10$ guarantees a breakout in such systems.

\section{Applicability to a shock driven by a wind}
\label{part: wind}
The solutions derived above are applicable to a strong (sudden) explosion. There are, however, situations in which energy is continuously injected
into the expanding ejecta as, e.g., in case of a continuous magnetar spin-down wind. Here we show that in many cases our solutions are applicable also for shocks driven by continuous energy injection. 

\begin{figure}
	\center
	\includegraphics[width=0.5\textwidth]{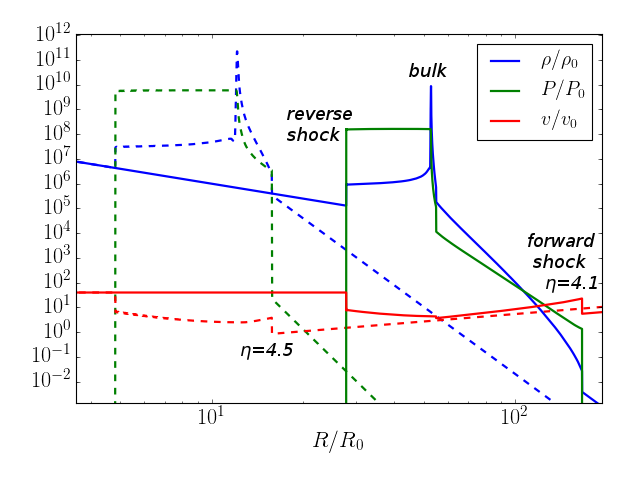}
	\caption{Two snapshots from a simulation of a shock driven by a continuous fast wind where the ejecta velocity profile is a broken power-law with indices $\alpha=9, \kappa=0$. At the time the first snapshot was taken (dashed lines), the forward shock is accelerating faster than the bulk of the ejecta, even though $\eta<\eta_c$. 
	In the second snapshot (solid lines), $\eta$ is smaller, and the shock is farther away from the bulk. The density, pressure, and velocity are plotted as a function of the radius. The forward shock, reverse shock, and bulk are indicated. Note that the bulk of the ejecta is located within the narrow density peak exhibited in the figure.}
	\label{fig:wind}
\end{figure}

Consider a fast wind driven into the center of a homologously expanding spherical ejecta with a density profile: 
\begin{equation}
\rho \propto
\begin{cases}
v^{-\kappa} & v<v_0, ~~\kappa<3\\
v^{-\alpha} & v\ge v_0, ~~\alpha>5
\end{cases} .
\end{equation}
The total mass and energy of the ejecta are carried almost entirely by the mass with $v<v_0$, such that $m(v<v_0) \approx M_{ej}$ and $E(v<v_0) \approx E_{ej}$.
At the contact between the wind and the ejecta, a forward shock and a reverse shock form; the forward shock traverse the ejecta, and the reverse shock traverse the wind. If the wind's energy injection rate is also a power-law, then as long as the forward shock is confined to the shallow density gradient region ($v<v_0$), the entire reverse-forward shock evolution has a self-similar solution \citep{Chevalier1984,Jun1998}. This solution breaks down at a time $t_0$, defined as the time at which the
forward shock reaches the beginning of the steep density gradient zone and starts accelerating. A snapshot from a simulation of such a setup at $t>t_0$ is shown in Fig. \ref{fig:wind}. As we show below, if the acceleration of the forward shock is fast enough, then it loses causal contact with the reverse shock and its evolution is similar to that of a shock driven by a strong explosion, namely, it propagates according to our analysis in the previous chapters. 

First, if the energy injection stops before $t_0$ then the reverse shock dies as the forward shock accelerates, so the solution is similar to that of an instantaneous explosion. If energy injection continues after $t_0$ then the evolution depends on the energy injection rate. As the forward shock is accelerating in the sharp density gradient, the deposited wind energy accelerates the bulk of the ejecta. The question is which of the two is accelerating faster. For simplicity, we will consider here only a constant wind energy injection rate, $L_w$, but the generalization to any type of injection rate is straightforward. If we assume that the wind is much faster than the bulk of the ejecta then the reverse shock is strong and almost all of the wind energy is deposited in the bulk of the ejecta, implying that the velocity of the bulk, $v_b$, satisfies  $\frac{1}{2}M_{ej} v_b^2 \approx L_w t$. Thus the bulk velocity evolves as 
\begin{equation}
v_b\propto t^{\frac{1}{2}}\propto r_b^{\frac{1}{3}}
\end{equation}
This equation shows that as long as $\delta>1/3$ the forward shock accelerates faster than the bulk and the solution we derived for a strong explosion is applicable. The initial value of $\delta$ at $t_0$ depends on $\alpha$ and $\eta(t_0) \approx \sqrt{L_wt_0/E_{ej}}$ and can be read from figure \ref{fig:delta(eta)}. It shows that accelerating shocks with $\alpha>w_c$ and $\eta(t_0)>\eta_c$ almost always have $\delta>1/3$ and since $\delta$ increases with time the forward shock will never regain causal contact with the bulk of the ejecta. In case of decelerating shocks, $\delta>1/3$ for even moderate values of $\eta$.  For example, $\eta \gtrsim 2$ is enough for $\alpha>8$ and $\eta \gtrsim 4$ is enough for $\alpha>6$. In such cases, the forward shock will get away from the bulk of the ejecta for some time until $\delta$ drops below $1/3$ and the bulk will start gaining over the shock until it will regain causal contact unless the wind stops or the forward shock breaks out before that.      

Finally, we note that here we considered a 1D spherically symmetric evolution that ignores the Rayleigh–Taylor instability that develops along the contact discontinuity. This instability does not affect the evolution of the forward shock unless it grows to the point that the Rayleigh–Taylor fingers reach to this shock. In this case, the shock may become highly non-spherical by the time that it gets to the steep density gradient. In that case, the shock acceleration during this phase enhances the a-sphericity \citep[e.g.,][]{Suzuki2017} and our spherical solution is not applicable.

\section{Summary}
\label{part: summary}
In this paper we study the propagation of a spherically symmetric shock in a steep power-law density gradient ($\rho_{ej} \propto t^{-3} v^{-\alpha}$, $\alpha>5$) of a homologously expanding medium. Such a shock may be driven by a strong explosion at the center of the ejecta or, in some circumstances, by a continuous fast wind that is driven by a central engine. 

We find that although such shocks are always accelerating in the lab frame, the fate of the shock depends on whether the ratio between the shock velocity and the ejecta velocity, $\eta$, increases or decreases along the shock trajectory.
We show that the shock solutions are divided into two families: growing shocks of increasing strength, where $\eta$ increases monotonically, that approach asymptotically the \citet{Waxman1993} solution for a shock in a static medium, and decaying shocks, where $\eta$ decreases, that weaken until they completely die out. The shock evolution depends on two parameters; the density power-law index $\alpha$ and the initial value of $\eta$. For  $\alpha<\omega_c$  ($\omega_c = 8.22\, [7.69]$ for $ \gamma=4/3\, [5/3] $) all the shocks  are decaying, while for $\alpha>\omega_c$ there is a critical value, $\eta=\eta_c(\alpha)$, that separates the branches of growing shocks and decaying shocks. If initially $\eta>\eta_c$ the shock accelerates relative to the ejecta and $\eta \to \infty$, while for $\eta<\eta_c$ the velocity ratio decreases monotonically and $\eta \to 1$. 
For $\eta=\eta_c(\alpha)$ we find an unstable, self-similar solution in which the ratio between the shock velocity and the velocity of the matter just upstream of the shock remains constant throughout the evolution. Thus, dynamics of shocks in an expanding medium is vastly different than the dynamics of shocks in a static medium, for which a converging mass profile (i.e., a density power-law index of $ \omega>3 $) is sufficient to guarantee that the shock will not decay.

We find $\eta_c(\alpha)$ by solving the self-similar solution semi-analytically and verify the qualitative behavior of the solutions described above using 1D hydrodynamical numerical simulations. While there is no analytic solution to the entire shock profile in the general case, we do find a full analytic description for the shock evolution $R(t)$. This description is found based on the numerical survey we conduct for the parameter space  $(\eta,\alpha)$, with $\alpha$ between 6 and 14 (the values expected for supernovae ejecta), and coincides with the simulations in this space.

One important feature that bears directly on observables of shocks in expanding media, is the modified density profile remaining far behind the shock. 
In particular, it is expected to govern the properties of the observed emission which diffuses out through this matter.
Examining the altered density profile, we find that in the case of decaying shocks the density gradient steepens after the shock passage. For growing shocks, the density gradient behind the shock becomes shallower than the initial density profile of the unshocked ejecta. Interestingly, all growing shock solutions we examined (with $\alpha$ between 9 and 14) produce a  rather similar post-shock density profile with a power-law index in the range $\sim 7-8$.

Finally, we examined the conditions for the shock to break out of the ejecta and produce a bright breakout signal. We find that for typical supernova ejecta, where the maximal velocity of the outflow is about a factor of $\sim 10$ larger than that of the bulk, an explosion with energy larger by a factor $\sim 3-10$ than the ejecta energy is sufficient for a successful shock breakout. 

\acknowledgments
We thank Nuriel Bitton for providing the code used to visualize the simulations.
This research was partially supported by the  Israel Science Foundation grant 1114/17 and by an ERC grant (JetNS). 

\appendix
\section{A. Derivation of the analytic solution}
\label{Appendix A}
Assuming the linear relation seen in figure \ref{fig:linear} holds, we denote $\eta-1=h$, and write the relation: 
\begin{equation}
    \frac{d\log(\dot{R}-v)}{d \log t}=ah - b.
    \label{dlogv_rel_app}
\end{equation}
recalling $\dot{R}= \eta v=\eta \frac{R}{t}$, we rewrite this equation: 
\begin{equation}
\begin{split}
    \frac{d\log(\dot{R}-v)}{d \log t}=\frac{d\log(\frac{R}{t}h)}{d \log t}&=\frac{d\log(\frac{R}{t})}{d \log t}+\frac{d\log(h)}{d \log t}\\
    &= h+\frac{d\log(h)}{d \log t}
    \end{split}
\end{equation}
plugging into (\ref{dlogv_rel_app}), we find: 
\begin{equation}
    \frac{d \log(h)}{d \log t}=(a-1)h - b.
\end{equation}
We can further simplify, $\frac{d\log h}{d\log t}=\frac{1}{h}\frac{d h}{d \log t} =\frac{1}{h}\frac{d \eta}{d \log t}$, and use this form to compare to Eq. (\ref{eq:deta_dt}). 
\begin{equation}
\frac{d \eta}{d\log t}=\eta\left( (\delta-1)\eta+1\right) = h\left((a-1)h-b\right) ~.
\label{dhdt}
\end{equation}
Rearanging this equations and rewriting it in terms of $h$ we obtain: 
\begin{equation}
\delta =\frac{h \left(a h-b+1\right)}{(h+1)^{2}}
\end{equation}
Examining the limit $h\to \infty, \delta\to\delta_{WS}$, we find that $a=\delta_{WS}$. 

Returning to equation \ref{dhdt}, we first require $\dot{h}$ to vanish for $h=\eta_c-1$, and find that $b = (\delta_{WS}-1)(\eta_c-1)$. For non vanishing $\dot{h}$, we may integrate:
\begin{equation}
\begin{split}
\log(\tilde{t}) &= \int \frac{dh}{h((\delta_{WS}-1)h-b)}\\
&=-\frac{1}{b} \log\left(\frac{h}{(\delta_{WS}-1)h-b}\right) +C
\end{split}
\end{equation}
where $\tilde{t}$ denotes the dimensionless time $\frac{t}{t_i}$, $t_i$ is the time at which the initial conditions are given. Requiring that at $\tilde{t}=1$, $h=h_i$, we find that $C=\log\left(\frac{(\delta_{WS}-1)h_i-b}{h_i}\right)$. Taking the exponent of both sides: 
\begin{equation}
\tilde{t}^{-b} = \frac{(\delta_{WS}-1) h_i -b}{h_i} \frac{h}{(\delta_{WS}-1) h -b}
\end{equation}or equivalently: 
\begin{equation}
    h=\frac{b h_i \tilde{t}^{-b}}{b - \sigma(\tilde{t})}
\end{equation}Where $\sigma(\tilde{t}) = \left(\delta_{WS}-1\right)h_i \left(1-\tilde{t}^{-b}\right)$ . We can now use this equation to find $R(t)$. Replacing $h=\frac{\dot{R}t}{R}-1$, 

\begin{equation}
\frac{\dot{R}\tilde{t}}{R}=1+\frac{bh_{i}\tilde{t}^{-b}}{b - \sigma(\tilde{t})}
\label{dot_Rt/R}
\end{equation}where we use $\dot{R}$ to denote the derivative of the shock velocity in terms of the dimensionless time;  $\dot{R} = \frac{d R}{d \tilde t}$. 
Integrating:
\begin{align}\label{eq:R_t}
\frac{R}{R_{i}}&=\exp \left(\intop_{1}^{\tilde{t}}\frac{1}{t'}
\frac{b(1+h_it^{'-b})-\sigma(t')}{b-\sigma(t')}dt'\right)\\
&=\begin{cases}
\tilde{t}\left(\frac{b}{b+\tilde{t}^{-b}\sigma(\tilde{t})
}\right)^{\frac{1}{\delta_{WS}-1}}, &\delta_{WS}\neq1\\ \nonumber
\tilde{t}\exp\left(\frac{\left(1-t^{-b}\right)h_{i}}{b}\right), &\delta_{WS}=1
\end{cases} 
\end{align}
Using equations \ref{dot_Rt/R} and \ref{eq:R_t}, we can now find $\dot{R}(t)$, 
\begin{align}
\dot{R} &=\left(1+\frac{bh_{i}\tilde{t}^{-b}}{b - \sigma(\tilde{t})}\right) \frac{R}{\tilde{t}} \\
&=R_i \left(1+\frac{bh_{i}\tilde{t}^{-b}}{b - \sigma(\tilde{t})}\right) 
\begin{cases}
\left(\frac{b}{b+\tilde{t}^{-b}\sigma(\tilde{t})}\right)^{\frac{1}{\delta_{WS}-1}}, &\delta_{WS}\neq1\\ \nonumber
\exp\left(\frac{\left(1-t^{-b}\right)h_{i}}{b}\right), &\delta_{WS}=1~.
\end{cases} 
\end{align}
This equation shows that for growing shocks where $h_i>\eta_c-1$ the initial conditions are such that $\frac{b}{\delta_{\text{WS}}-1}<h_i$ and as $t\to\left(1-\frac{b}{h_{0}\left(\delta_{\text{WS}}-1\right)}\right)^{-\frac{1}{b}}$, the denominator in the first term approaches zero, and the shock velocity diverges. Thus in growing shocks the shock velocity diverges at a finite time.
If $\delta_{WS}\le 1$ or $\delta_{WS}>1$ and $\frac{b}{\delta_{\text{WS}}-1}>h_{0}$, then the shock is decaying as evident from the fact that the denominator does not vanish, and:
\begin{equation}
\lim_{t\to \infty}\dot{R}
=R_{i}\begin{cases}
\left(\frac{b}{b-\left(\delta_{\text{WS}}-1\right)h_{i}}\right)^{\frac{1}{\delta_{WS}-1}}, & \delta_{WS}\neq1\\
\exp\left(\frac{h_{i}}{b}\right), & \delta_{WS}=1,
\end{cases}
\end{equation}
which marks the velocity of the shell that the shock never cross.

Using the results above, we can derive the expression for the density profile far behind the shock. First, 
\begin{align}
\frac{d \log \eta}{d \log R} &= \frac{d\log\eta}{d\log t}\frac{d\log t}{d\log R}\\ &= \frac{1}{\eta^{2}}\frac{d\eta}{d\log t}\nonumber \\ &=\frac{h}{\left(h+1\right)^{2}}\left(\left(\delta_{WS}-1\right)h+b\right) \nonumber
\end{align}
Now plugging the result and the expression for $\delta$ into \ref{eq:rho_f_exp}:
 \begin{align}
 \frac{d \log \rho_f}{d \log v_f} &= -\alpha+\frac{\alpha}{\delta}\frac{d \log \eta}{d \log R}\\
 &=-\alpha\frac{h+1}{1-b+h\delta_{\text{WS}}}\nonumber
 \end{align}
 
\clearpage
\bibliography{Spherical_shocks}
\bibliographystyle{mnras}

\end{document}